\documentstyle[12pt,epsfig]{article}

\newcommand{\beq}{\begin{equation}}
\newcommand{\eeq}{\end{equation}}
\newcommand{\bea}{\begin{eqnarray}}
\newcommand{\eea}{\end{eqnarray}}
\def\maty   {|\overline{\cal{M}}_2|^2}
\def\matz   {|\overline{\cal{M}}_3|^2}

\def\lapprox{\lower .7ex\hbox{$\;\stackrel{\textstyle <}{\sim}\;$}}
\def\gapprox{\lower .7ex\hbox{$\;\stackrel{\textstyle >}{\sim}\;$}}

\begin{document}
\titlepage

\begin{flushright}
{CERN-TH/98-150}\\
{DTP/97/104}\\
May 1998
\end{flushright}

\begin{center}
\vspace*{2cm}
{\Large {\bf Topology of Hadronic Flow for Higgs Production \\[2mm]
at Hadron Colliders}}\\

\vspace*{1.5cm}
M.\ Heyssler$^a$, V.\ A.\ Khoze$^{a,b,c}$ and W.\ J.\ Stirling$^{a,d}$ \\

\vspace*{1.0cm}
$^a  $ {\it Department of Physics, University of Durham,
Durham DH1 3LE, England }\\[2mm]

$^b  $ {\it INFN - Laboratori Nazionali di Frascati, I-00044 Frascati (Roma),
 Italy}\\[2mm]

$^c  $ {\it TH Division, CERN, CH-1211 Geneva 23, Switzerland}\\[2mm]

$^d  $ {\it Department of Mathematical Sciences, University of Durham,
Durham DH1 3LE, England }
\end{center}

\vspace*{2cm}
\begin{abstract}
Hadronic radiation  provides a tool to
distinguish different topologies of colour flow in
hard scattering processes. We study the structure of hadronic flow corresponding
to Higgs production and decay in high--energy hadron--hadron collisions.
In particular, the signal $gg \to H \to b \bar b$ and background
$gg \to b \bar b$ processes are shown to have very different
radiation patterns, and this may provide an useful additional
 method for distinguishing Higgs signal events from the QCD background.
\end{abstract}
\newpage

\setcounter{page}{1}
\pagestyle{plain}


\section{Introduction}

The distribution of soft hadrons or jets accompanying energetic
final--state particles in hard scattering processes
is governed by the underlying colour dynamics at short
distances [\ref{ref:DKT}--\ref{ref:emw}] . The soft hadrons paint the colour 
portrait of the parton hard scattering,
and can therefore act as a `partonometer' [\ref{ref:DKT}--\ref{ref:klo}]. 
Since signal and background
processes at hadron colliders can have very different colour structures
(compare for example the $s$--channel 
colour singlet process $q \bar q \to Z' \to q' \bar q'$
 with the colour octet process
$q \bar q \to g^* \to q' \bar q'$), the distribution of accompanying
soft hadronic radiation in the events can provide a useful additional
diagnostic tool for identifying new physics processes.

Quite remarkably, because of the property of Local Parton Hadron Duality
(see for example Refs.~\cite{DKMT,book,adkt}) the distribution of soft
 hadrons can be well described by the
amplitudes for producing a single additional soft gluon.
This distribution
 takes the form of a soft `antenna pattern' distribution multiplying the 
leading--order hard scattering matrix element squared. Confirmation of the validity
of this approach comes from recent studies of the production of 
soft hadrons and jets
accompanying large $E_T$ jet and $W+$jet production  by the CDF
\cite{CDF}  and
D0 collaborations  \cite{D0} at the Fermilab Tevatron.

One of the most important physics goals of the CERN LHC $pp$ collider is
the discovery of the Higgs boson \cite{LHC}. Many  scenarios, corresponding
to different production and decay channels, have been investigated, see
for example the studies reported in Refs.~\cite{ATLAS,CMS}. While
final states containing leptons and photons are relatively background free,
they generally have very small branching ratios. In contrast, the more
probable decay channels involving (heavy) quarks have large QCD backgrounds.
The question naturally arises whether hadronic radiation patterns 
could help distinguish such signals from backgrounds. We have in mind the 
following type of scenario. Suppose an invariant mass peak is 
observed in a sample of (tagged) $b \bar b$ events. If these correspond
to Higgs production, then the distribution of accompanying soft radiation
in the event\footnote{We take this to mean the angular 
distribution of hadrons or  `minijets' with 
energies of at most a few GeV, well separated from the beam and final--state
energetic jet directions.} 
will look very different from that expected in background
QCD production of $b \bar b$ pairs. One could imagine, for example, comparing the
topologies of the hadronic flows `on and off resonance'. 

In this study we will consider the hadronic radiation patterns for
two of the standard Higgs processes at LHC: direct production
 $ gg \to H \to b \bar b$
and associated production $ q \bar q ' \to W H \to \ell \nu_\ell b \bar b$.
Although the non-zero $b$--quark mass is largely irrelevant when 
computing the radiation patterns, we will also consider the case when the
final--state quark mass is large, so that our analysis can also be applied
for example to $ H \to t \bar t$.
Our Higgs analysis is a natural extension of the  
studies of Refs.~\cite{Ell96,Hey97}, where the antenna patterns for $Z'$ 
in $p \bar p$ collisions and 
leptoquark production in $ep$ collisions 
were calculated and shown to be different from those
of the QCD backgrounds.

The analysis presented here should be regarded as a `first look' at the 
possibilities offered by hadronic flow patterns in searching for the Higgs.
Of course, ultimately there is no substitute for a detailed Monte Carlo
study including detector effects. However the results presented here 
indicate that the effects can be potentially  large, and therefore
that more detailed studies are definitely worthwhile.

The paper is organised as follows. In the following section we consider
direct production and $q\bar q$ decay of the Higgs boson, first for massless
and  then for massive quarks. Section~3 extends the analysis to associated
production and Section~4 presents our conclusions.

\section{Hadronic radiation patterns for signal and
background processes}
\label{sec:antenna}

\begin{figure}[hbt] 
\unitlength1cm 
\begin{center} 
\epsfig{figure=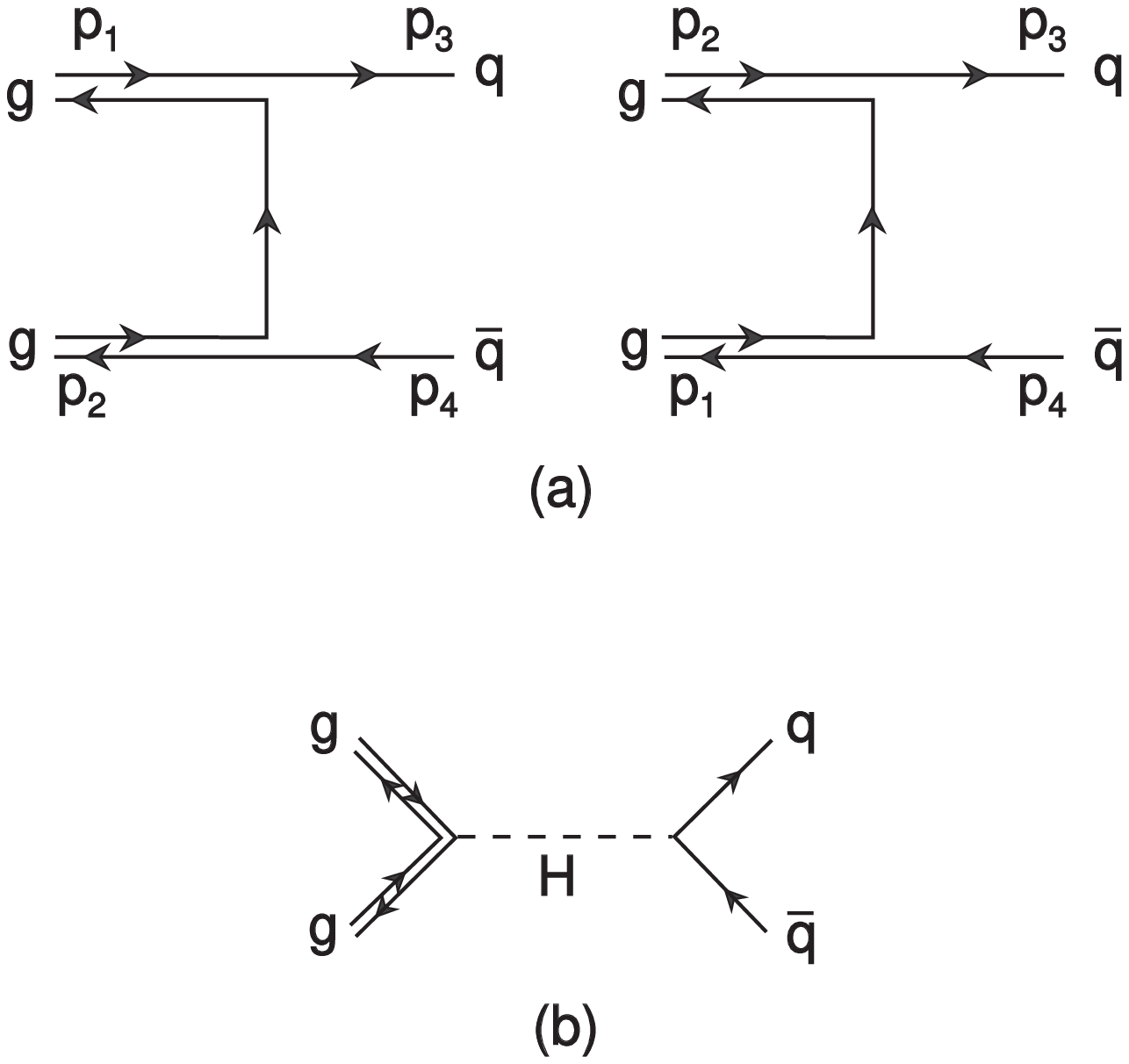,width=9cm}
\end{center}
\caption[]{The  colour flow  diagrams  for  the   processes  (a) $gg\rightarrow
q\bar{q}$ and (b) $gg\rightarrow H \rightarrow q\bar{q}$.}
\label{fig:flow}
\end{figure}

We begin by considering the hadronic radiation patterns for the
signal ($gg \rightarrow H \rightarrow q\bar{q}+g$)
and background ($gg \rightarrow q\bar{q}+g$)
production of a massless $q \bar q$ pair. The
impact of non--zero quark masses will be considered later.
The radiation pattern is defined as the ratio of the $2 \to 3$
and $2 \to 2$ matrix elements using the soft--gluon approximation
for the former. The dependence on the soft gluon momentum $k$ then enters
via the eikonal factors (`antennae') \cite{Azi85}
\beq \label{eq:eikonal}
\lbrack ij\rbrack = \frac{p_ip_j}{(p_ik)(p_jk)}\; .
\eeq 
For the QCD background process 
$g(p_1)g(p_2)\rightarrow   q(p_3)\bar{q}(p_4)+g(k) $ we have
\begin{eqnarray} \label{eq:ggqqg} \nonumber
\frac{1}{g_s^6}  \matz  (gg\rightarrow   q\bar{q}+g)  &=&  \frac{1}{2}
(t^2+u^2)      \left[\left(1-\frac{1}{N_c^2}\right)\frac{1}{tu}      -
\frac{2}{s^2}\right]   \left\{    \frac{N_c}{C_F}\lbrack12\rbrack    +
\lbrack34\rbrack   \right\}  \\  \nonumber  &-&   \frac{1}{8}(t^2+u^2)
\left[\left(1-\frac{2}{N_c^2}\right)\frac{1}{tu}                     -
\frac{2}{s^2}\right]    \left\{     \frac{N_c}{C_F}\lbrack12;34\rbrack
\right\}   \\   &+&    \frac{1}{8}(t^2-u^2)    \left[\frac{1}{tu}    -
\frac{2}{s^2}\right]    \left\{   \frac{N_c}{C_F}    (\lbrack14\rbrack
+\lbrack23\rbrack - \lbrack13\rbrack - \lbrack24\rbrack )\right\}\; ,
\end{eqnarray}
with $s=(p_1+p_2)^2,\; t=(p_1-p_3)^2,\; u=(p_1-p_4)^2$, and
\beq \label{eq:deikonal} 
\lbrack  ij;kl  \rbrack = 2\lbrack ij \rbrack + 2\lbrack  kl \rbrack -
\lbrack ik \rbrack  -\lbrack il \rbrack - \lbrack jk \rbrack - \lbrack
jl \rbrack \; .
\eeq 
This is to be normalised by the matrix element for the
leading--order scattering process $g(p_1)g(p_2)\rightarrow q(p_3)\bar{q}
(p_4)$:
\beq \label{eq:ggqq} 
\frac{1}{g_s^4} \maty (gg\rightarrow q\bar{q})  = \frac{1}{2}(t^2+u^2)
\left[\frac{1}{N_c}\frac{1}{tu} - \frac{1}{C_F}\frac{1}{s^2}\right]\; .
\eeq 
The antenna pattern is then
\beq \label{eq:qcdrat}
{\cal{R}}^{QCD}  =  g_s^{-2}\frac{\matz   (gg\rightarrow  q\bar{q}+g)}
{\maty (gg\rightarrow q\bar{q})}\; .
\eeq
Note that 
because of the non--trivial colour structure of the leading--order
Feynman diagrams, see Fig.~\ref{fig:flow}(a),
there is no simple factorisation of the eikonal factors.
This is in contrast to the signal (Higgs) process, for which
\beq \label{eq:higgsrat}
{\cal{R}}^H   =    g_s^{-2}\frac{\matz    (gg\stackrel{H}{\rightarrow}
q\bar{q}+g)}   {\maty   (gg\stackrel{H}{\rightarrow}    q\bar{q})}   =
2N_c\lbrack12\rbrack + 2C_F\lbrack34\rbrack \; ,
\eeq
with the same momentum labeling.
The two terms correspond to gluon radiation off the initial
state gluons (colour factor $N_c$) and the final--state
quarks (colour factor $C_F$). With colour--singlet exchange in the
$s$--channel (Fig.~\ref{fig:flow}(b)),
there is no interference between the initial-- and final--state 
emission, in contrast to the QCD background antenna pattern.
It is this feature which will give rise to significant quantitative
differences between the radiation patterns (see below).

The next step is to define the kinematics.
The four momenta are labelled by
\begin{equation}
a(p_1) + b(p_2)\to c(p_3) + d (p_4) + g(k)\, ,
\label{eq:generic}
\end{equation}
where the  gluon is assumed soft relative to the two
large-$E_T$ partons $c$ and $d$, i.e. $k \ll E_T$.
 Ignoring the gluon momentum in the
energy-momentum constraints, working in the subprocess centre--of--mass
frame, and
using the notation $p^\mu = (E,p_x,p_y,p_z)$, we have
\begin{eqnarray}
p_1^\mu &=& (E_T\cosh\eta, 0, 0, E_T\cosh\eta) \;  , \nonumber \\
p_2^\mu &=& (E_T\cosh\eta, 0, 0, -E_T\cosh\eta) \;  , \nonumber \\
p_3^\mu &=& (E_T\cosh\eta, 0, E_T, E_T\sinh\eta) \;  , \nonumber \\
p_4^\mu &=& (E_T\cosh\eta, 0, -E_T, -E_T\sinh\eta) \;  , \nonumber \\
k^\mu &=& (k_T\cosh(\eta+\Delta\eta), k_T\sin\Delta\phi,
 k_T \cos\Delta\phi, k_T\sinh(\eta+\Delta\eta)) \;  .
\label{eq:kin2to2}
\end{eqnarray}
This is the appropriate form for studying the angular distribution of the soft
gluon jet relative to the large--$E_T$ jet 3, the separation between
these being parametrised by $\Delta\eta$ and $\Delta\phi$.
In terms of these variables, the soft gluon phase space is
\begin{equation}
{1\over (2 \pi)^3} \; {d^3 k \over  2 E_k}
 = {1 \over 16\pi^3}\; k_T d k_T \; d \Delta\eta\;  d \Delta\phi\, .
\end{equation}
We will be particularly interested in the shape of the radiation pattern
as a function of the variables $\Delta\eta$ and $\Delta\phi$.
Note that the direction of the soft gluon is measured with respect to
the $p_3$ jet. Thus for massless $2\to 2$ scattering, collinear singularities
are located at $ \Delta\eta = 0,\; \Delta\phi=0$ and
$ \Delta\eta = -2\eta,\; \Delta\phi=\pi$.
\begin{figure}[htb] 
\unitlength1cm 
\begin{center} 
\epsfig{figure=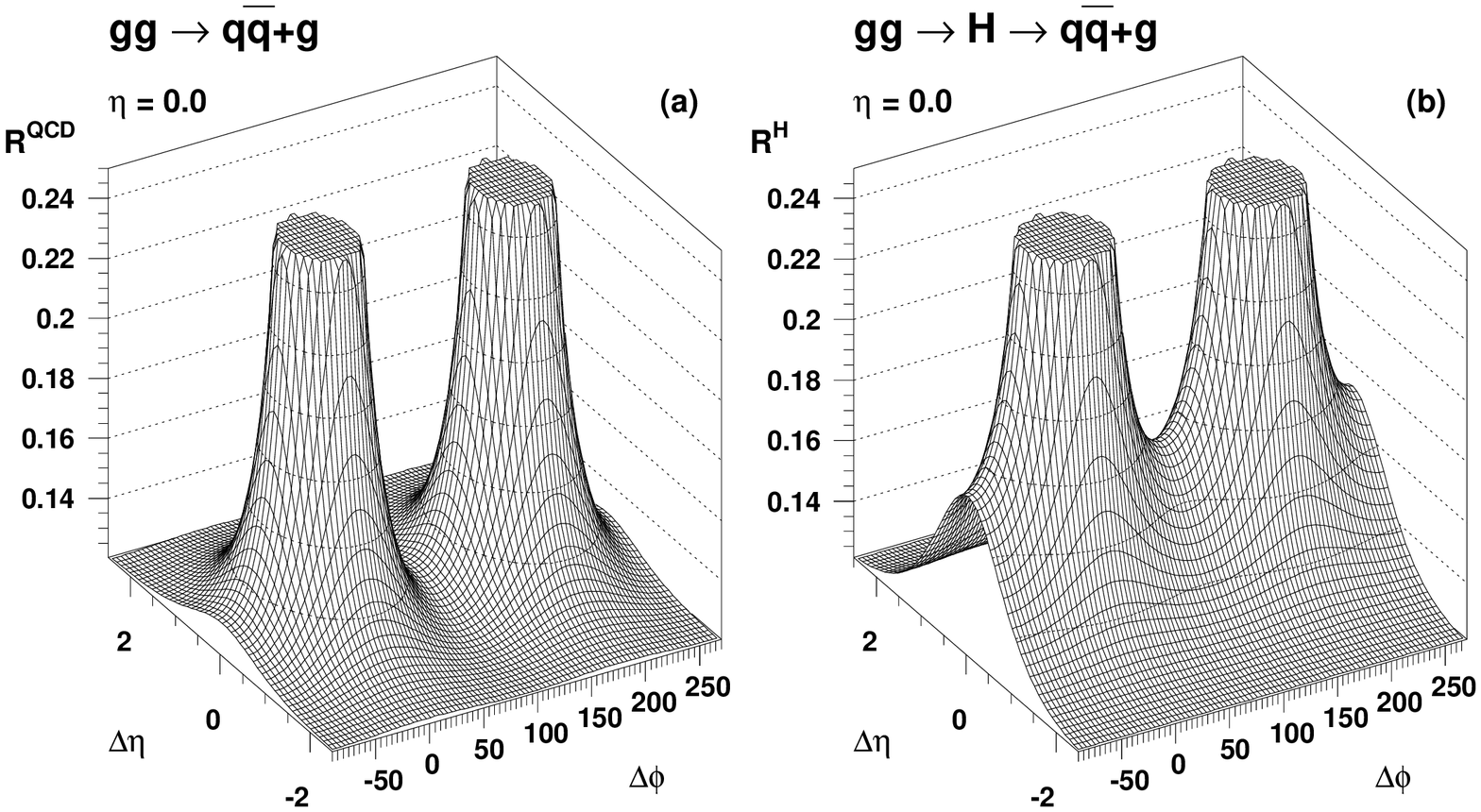,width=\textwidth}
\end{center}
\caption[]{The      antenna     patterns      ${\cal{R}}^{QCD}$     of
Eq.~(\ref{eq:qcdrat})    and    ${\cal{R}}^H=2N_c\lbrack12\rbrack    +
2C_F\lbrack34\rbrack$  of  Eq.~(\ref{eq:higgsrat})  for the  processes
$gg\rightarrow    q\bar{q}+g$   and   $gg\rightarrow   H   \rightarrow
q\bar{q}+g$, with $\eta = 0$ and $k_T = 10$~GeV. The units of ${\cal R}$
are GeV$^{-2}$.}
\label{fig:eta0}
\end{figure}

\begin{figure}[htb] 
\unitlength1cm 
\begin{center} 
\epsfig{figure=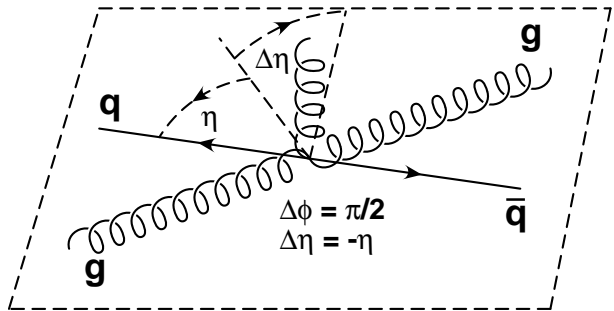,width=8cm}
\end{center}
\caption[]{Sketch of the kinematics for the case of observing the soft
gluon   perpendicular   to  the   event   plane   at the `symmetric point'
${\cal{P}}_{\rm c} = (\Delta\eta= -\eta, \Delta\phi=\pi/2)$.}
\label{fig:event}
\end{figure}

We first study the QCD and Higgs radiation patterns
 for central $q \bar q$ jets, i.e. $\eta = 0$. Using the kinematics
of  Eq.~(\ref{eq:kin2to2}) with $\eta=0$, Eq.~(\ref{eq:higgsrat}) gives
\beq \label{eq:higgsrat0}
{\cal{R}}^H|_{\eta=0}   =  \frac{4}{k_T^2}   \frac{N_c   \left(\cosh^2
(\Delta \eta) -  \cos^2(\Delta\phi)\right) + C_F} {\cosh^2(\Delta\eta)
- \cos^2(\Delta\phi)}  \; ,
\eeq
and
\begin{eqnarray} \label{eq:qcd0} \nonumber
{\cal{R}}^{QCD}|_{\eta=0}  &=&  \frac{2}{k_T^2}  \frac{N_c^2  \left( 2
\cosh^2(  \Delta\eta) - \cos^2  (\Delta \phi) -1 \right)  }{\left\{
\left( 4C_F-N_c  \right) \left(  \cosh^2(\Delta\eta)  - \cos^2( \Delta
\phi) \right) \right\} }\\ &+& \frac{4}{k_T^2}  \frac{2 N_c \left( 1 -
\cosh^2(\Delta\eta)  \right) + C_F  \left(  N_c^2 - 2  \right)  } {N_c
\left \{ \left( 4 C_F - N_c \right)  \left(  \cosh^2(  \Delta \eta ) -
\cos^2( \Delta \phi ) \right) \right\} }\; .
\end{eqnarray}
Note that the radiation  patterns are independent  of $E_T$.
Fig.~\ref{fig:eta0}  shows the dependence of ${\cal{R}}^{H}$ and
${\cal{R}}^{QCD}$ on $\Delta\eta$ and $\Delta\phi$.
It is straightforward to show that the patterns are identical
close to the beam direction,
\beq \label{eq:inflim}
\lim_{|\Delta\eta|\rightarrow\infty}
{\cal{R}}^{H,QCD} =    \frac{4}{k_T^2} N_c \; ,
\eeq
independent of $\Delta\phi$,
and close to the directions of the final--state quarks,
\beq \label{eq:collim}
\lim_{\Delta\eta,\Delta\phi \rightarrow 0}
{\cal{R}}^{H,QCD} \to   \frac{4C_F}{k_T^2} \frac{1}{\cosh^2(\Delta\eta) -
\cos^2(\Delta\phi)}\; .
\eeq
The main difference arises from the amount of radiation
{\it between} the final-state quark jets. To study this further we
consider the distributions at the symmetric point ${\cal{P}}_{\rm c}$
located at $(\Delta\eta = - \eta = 0, \Delta\phi = \pi/2)$.
This corresponds to soft gluon radiation perpendicular to the
plane of the $gg\to q \bar q$ scattering, see Fig.~\ref{fig:event}.
Again  using  the kinematics of  Eq.~(\ref{eq:kin2to2}),  we find for
the QCD background process
\beq \label{eq:qcd00pi2}
{\cal{R}}^{QCD}|_{\eta=0}({\cal{P}}_{\rm c}) =
\frac{2}{k_T^2}\frac{2C_F(N_c^2-2)+N_c^3}{N_c(4C_F-N_c)} \sim 0.1304\; ,
\eeq
where the numerical value corresponds to   $N_c   =3$   and
$k_T=10$~GeV.
In contrast, for the Higgs signal process we find
\beq \label{eq:higgs00pi2}
{\cal{R}}^H|_{\eta=0}({\cal{P}}_{\rm c}) =
\frac{4}{k_T^2}\left( C_F + N_c\right) \sim 0.1733\; .
\eeq
There is therefore approximately $4/3$ more radiation between the final--state
 jets for the Higgs production process. This is due to the absence of a colour
string connecting the final--state quarks in the QCD background process,
see Fig.~\ref{fig:flow}(a).

\begin{figure}[htb] 
\unitlength1cm 
\begin{center} 
\epsfig{figure=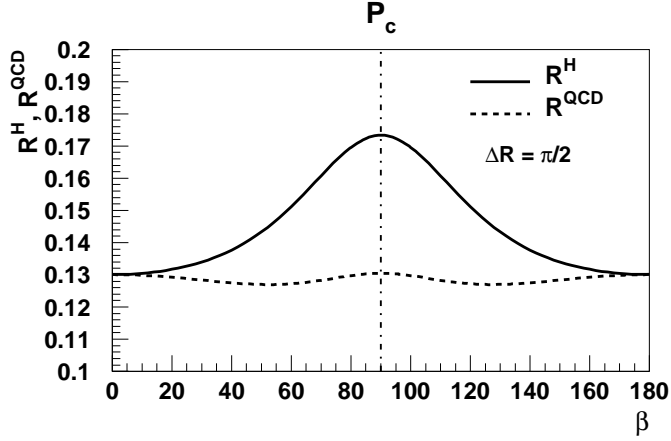,width=10cm}
\end{center}
\caption[]{The dependence  of the
antenna patterns ${\cal{R}}^{H}$ and ${\cal{R}}^{QCD}$ on
the angular variable $\beta$ defined in Eq.~(\protect\ref{eq:legovar}).
The units of ${\cal R}$
are GeV$^{-2}$.}
\label{fig:beta}
\end{figure}

The QCD background process does, however, have colour strings
connecting the initial-- and final--state quarks, and this leads
to an enhancement of soft radiation between the jets
 in the plane of the scattering. We can illustrate this
by considering the radiation patterns around the direction
of the final state quark. In particular we introduce (as in
\cite{Ell96}) the variables $(\Delta R,\beta)$, where
\beq
\label{eq:legovar}
\Delta \eta = \Delta R \cos\beta\; , \quad
\Delta \phi = \Delta R \sin\beta\; .
\eeq
For fixed $\Delta R > 0$, varying $\beta$ between $0$ and $2\pi$
describes a circle in the $(\Delta\eta,\Delta\phi)$ plane around
the quark direction. In addition, if we fix $\Delta R  = \pi/2$
then the symmetric point ${\cal{P}}_{\rm c}$ corresponds to
$\beta  = \pi/2$ (or equivalently $3\pi/2$), and the soft gluon
is in the $2\to 2$ scattering plane for $\beta = 0, \pi$.
Figure~\ref{fig:beta} shows the dependence of the radiation patterns
${\cal{R}}^{H,QCD}$ on $\beta$ for $\Delta R  = \pi/2$, as before
for $\eta = 0$ final--state quarks. At $\beta = \pi/2$ we
have ${\cal{R}}^{H} > {\cal{R}}^{QCD}$, as discussed above,
whereas  at $\beta = 0, \pi$ we have
  ${\cal{R}}^{QCD} = {\cal{R}}^{H}$.\footnote{In fact, the equality
  of the distributions at $\beta = 0,\pi$ is true for all $\Delta R$.}
The {\it shape} of the $\beta$ distribution therefore provides
a powerful discriminator between signal and background.

\begin{figure}[ht] 
\unitlength1cm 
\begin{center} 
\epsfig{figure=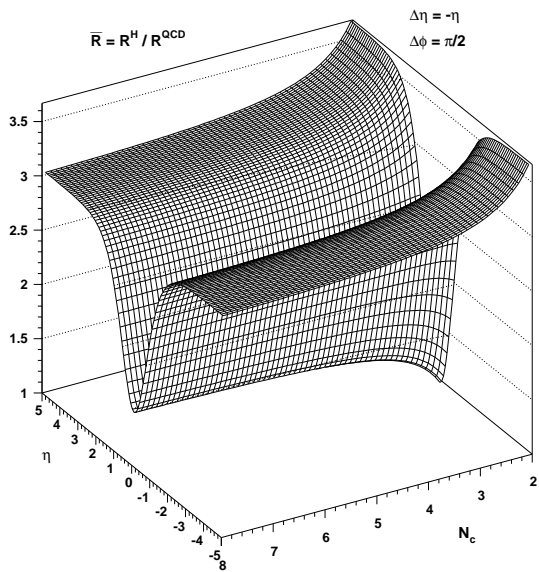,width=8.0cm}
\end{center}
\caption[]{The          ratio           $\overline{\cal{R}}          =
{\cal{R}}^H/{\cal{R}}^{QCD}$  as a  function  of $N_c$ and  $\eta$
at the symmetric interjet point
$\Delta\eta = -\eta$ and $\Delta\phi = \pi/2$.}
\label{fig:nceta}
\end{figure}

How does the interjet radiation
enhancement depend on the jet rapidity $\eta$?
Again we consider the symmetric point located  at  ${\cal{P}}_{\rm  c} =
(\Delta\eta=-\eta, \Delta\phi=\pi/2)$.
At this point  ${\cal{R}}^H$  is completely
independent of $\eta$,
\beq \label{eq:higgstriv}
{\cal{R}}^H({\cal{P}}_{\rm c})   =  \frac{4}{k_T^2}
\left( N_c + C_F \right)\; .
\eeq
which follows immediately from  Eq.~(\ref{eq:higgsrat}) since
$\lbrack 12 \rbrack = \lbrack 34 \rbrack = 2/k_T^2$ at ${\cal{P}}_{\rm
c}$.  The result is slightly  more  complicated  for  ${\cal{R}}^{QCD}$.
Here we find
\beq \label{eq:qcdnotriv}
{\cal{R}}^{QCD}({\cal{P}}_{\rm c})   =   \frac{2}
{k_T^2} \frac{1}{N_c} \frac{4C_F\cosh^2(\eta) \left( N_c^2 - 1 \right)
+ N_c^2 \left(N_c - 2C_F \right)}{ 4C_F\cosh^2(\eta) - N_c }\; .
\eeq
At ${\cal{P}}_{\rm c}$, ${\cal{R}}^{QCD}$ is maximal for $\eta=0$
with the  value given in Eq.~(\ref{eq:qcd00pi2}).  As  $|\eta|  \rightarrow
\infty$  ${\cal{R}}^{QCD}$ approaches its minimum value,
\beq \label{eq:qcdinf}
\lim_{|\eta|\rightarrow\infty} {\cal{R}}^{QCD}({\cal{P}}_{\rm c})
= \frac{4} {k_T^2} C_F\; .
\eeq
Note that in the  large--$\eta$  limit  the  ratio
$\overline{\cal{R}}\equiv {\cal{R}}^{H}/{\cal{R}}^{QCD}$
 at ${\cal{P}}_{\rm c}$ is significantly larger than its value
at $\eta = 0$:
\begin{eqnarray}
{\overline{\cal{R}}}(\eta=0, {\cal{P}}_{\rm c} )
 & = &  \frac{3N_c^4 - 7N_c^2 + 2}{2N_c^4 - 3N_c^2 + 2}  =
1.3285\; , \nonumber \\
{\overline{\cal{R}}}(|\eta|\to\infty, {\cal{P}}_{\rm c} )
 & = &  \frac{3N_c^2 - 1}{N_c^2 - 1} = 3.25\; .
\label{eq:maxrat0}
\end{eqnarray}
In other words, the difference in the signal and background radiation
patterns at the symmetric interjet point increases with increasing jet
rapidities. Note that the large--$N_c$ limits of the
ratios in Eq.~(\ref{eq:maxrat0}) are simply $3/2$ and $3$, and also that
${\overline{\cal{R}}}(\eta=0, {\cal{P}}_{\rm c} ) = 1$ for $N_c = 2$.
This is illustrated in Fig.~\ref{fig:nceta} which shows the dependence
of ${\overline{\cal{R}}}$ evaluated at ${\cal{P}}_{\rm c}$
 on $\eta$ and $N_c$.
%


\subsection{Massive quarks}

So far we have  only  considered  massless  quarks. In fact for
$H\to b \bar b$, with $m_b \ll M_H$, this should be an excellent
approximation, since the soft gluon only `feels' the finite $b$--quark
mass very close to the jet axis, where our analysis does not 
in any case apply.
Far from the jet direction, and in particular at the symmetric point
${\cal{P}}_{\rm c}$, the effect of the non--zero $b$ mass will
be negligible.
The situation is however very different for the case of $H \to t \bar t$,
at $M_H \gapprox 2 m_t$. Now mass effects {\it are} important
in the radiation pattern, as we shall demonstrate below.

If we allow a  finite  mass  for the  produced
quarks then the  kinematics  have to be  changed  accordingly. Thus we
replace the kinematics of Eq.~(\ref{eq:kin2to2}) by
\begin{eqnarray}
p_1^\mu &=& (E_q, 0, 0, E_q) \; , \nonumber \\
p_2^\mu &=& (E_q, 0, 0, -E_q) \;  , \nonumber \\
p_3^\mu &=& (E_q, 0, p_T, E_q\tanh\eta)\;   , \nonumber \\
p_4^\mu &=& (E_q, 0, -p_T, -E_q\tanh\eta) \;  , \nonumber \\
k^\mu &=& (k_T\cosh(\eta+\Delta\eta), k_T\sin\Delta\phi,
 k_T \cos\Delta\phi, k_T\sinh(\eta+\Delta\eta))\;  ,
\label{eq:fourmass}
\end{eqnarray}
i.e. we denote the energy of the quark jets by
$E_q$ and their transverse momentum
by $p_T$. Thus
\beq \label{mqf}
E_q = \cosh(\eta) \sqrt{m_q^2 + p_T^2} \; .
\eeq
 We again work in the subprocess centre--of--mass frame.
It is convenient to introduce
the dimensionless variable $\Theta$ as the ratio of the final--state
quark mass $m_q$ to its energy
\beq \label{eq:index}
\Theta = \frac{m_q}{E_q}\; .
\eeq

For non--zero $m_q$ the antenna  patterns receive additional
contributions.  For  example,  the
antenna pattern of ${\cal{R}}^H$  of  Eq.~(\ref{eq:higgsrat}) becomes
\beq \label{massh}
{\cal{R}}^H_{\Theta} = {\cal{R}}^H  -C_F\lbrack 33 \rbrack -C_F\lbrack
44 \rbrack\; , 
\eeq
where   the
massive  equivalents of ${\cal{R}}^H$  and  ${\cal{R}}^{QCD}$
are labelled with the suffix $\Theta$.
One effect of the additional terms is to
cancel the final--state collinear singularities, leading instead to the
well--known  {\em  dead  cone}  \cite{Dok91a} phenomenon.
Using the results of  Ref.~\cite{Kho94},  we obtain a  somewhat  more
complicated  expression  for the  massive  equivalent  to
${\cal{R}}^{QCD}$,
\begin{eqnarray} \label{massq} \nonumber
{\cal{R}}^{QCD}_{\Theta}  &=&  \left(2N_c  -2C_F + 2{\cal{Y}}  \right)
\lbrack 12 \rbrack + \left(C_F - {\cal{X}} - {\cal{Y}} \right) \left\{
\lbrack 13 \rbrack + \lbrack 24 \rbrack  \right\} \\ &+&  \left(C_F  +
{\cal{X}} - {\cal{Y}}  \right) \left\{ \lbrack 14 \rbrack + \lbrack 23
\rbrack  \right\} +  2{\cal{Y}}  \lbrack 34  \rbrack  -C_F  \lbrack 33
\rbrack -C_F \lbrack 44 \rbrack\; ,
\end{eqnarray}
with
\begin{eqnarray} \nonumber \label{xmas}
{\cal{X}} &=& \frac{N_c^2}{4C_F}  \left[ (1+2\mu) \left( \frac{1}{U} -
\frac{1}{T}  \right)  - \mu^2  \left(  \frac{1}{U^2}  -  \frac{1}{T^2}
\right) + 2(U-T)  \right] \\ &\times&  \left[  \frac{1}{UT}-\frac{N_c}{C_F}
\right]^{-1}\; \left[ T^2 + U^2 + 2\mu - \frac{\mu^2}{UT} \right]^{-1}\; ,
\end{eqnarray}
and
\beq \label{ymas}
{\cal{Y}} = \frac{1}{4C_F} \left[  \frac{1}{N_c^2UT}+2\right]  \left[
\frac{1}{UT} - \frac{N_c}{C_F} \right]^{-1}\; .
\eeq
The variables $T$, $U$ and $\mu$ are defined as
\beq \label{vary}
T =  \frac{p_1p_3}{p_1p_2}\; ,  \qquad U = \frac{p_1p_4}{p_1p_2}\; ,  \qquad
\mu = \frac{m_q^2}{p_1p_2}\; .
\eeq
It is straightforward to show that the massless results are recovered
in the limit $m_q(\Theta) \to 0$.

\subsection{Threshold behaviour ($\Theta=1$)}

\begin{figure}[p] 
\unitlength1cm 
\begin{center} 
\epsfig{figure=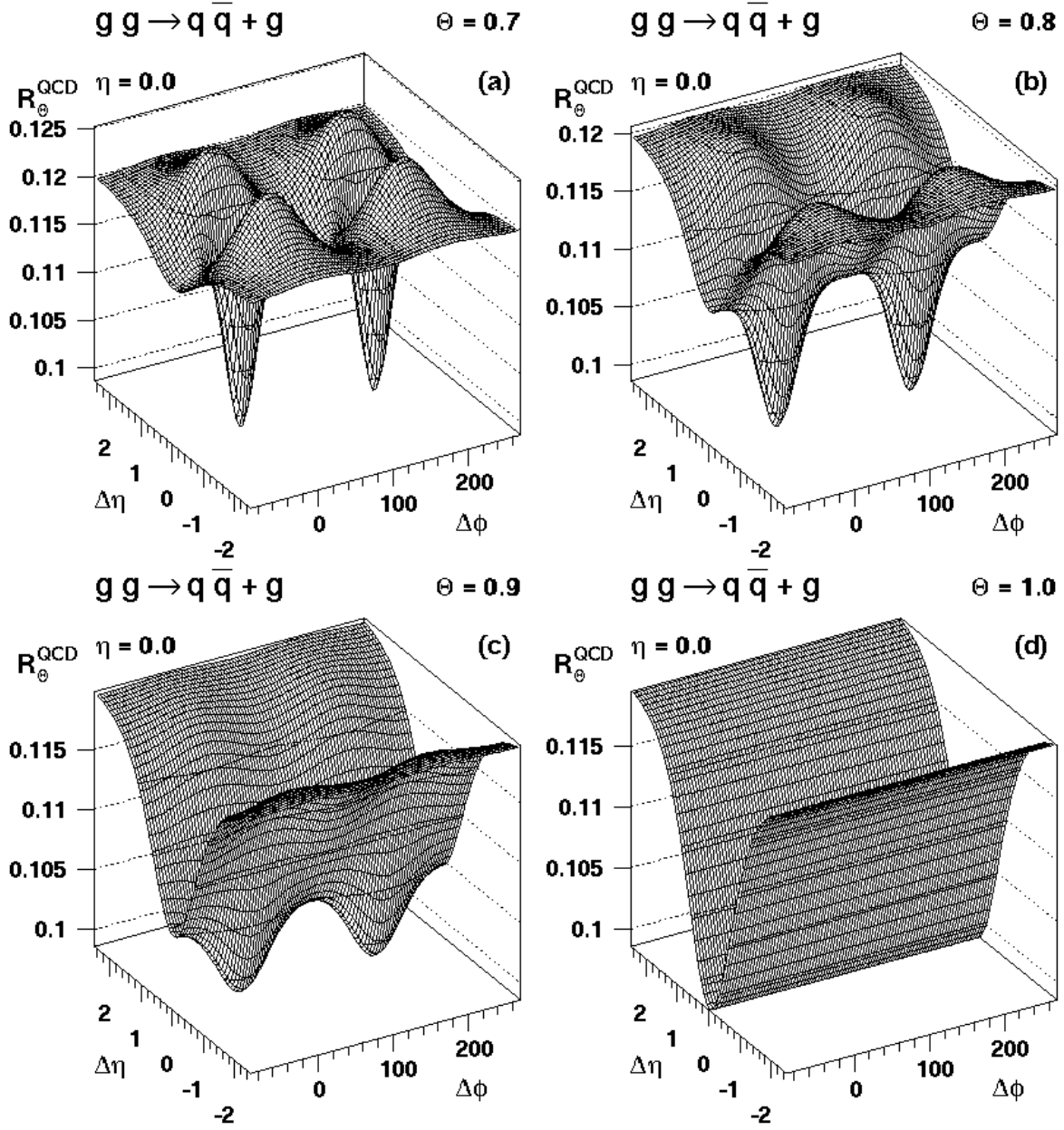,width=\textwidth}
\end{center}
\caption[]{The   antenna   patterns    ${\cal{R}}^{QCD}_{\Theta}$   of
Eq.~(\ref{massq})  for the process  $gg\rightarrow  q\bar{q}+g$  with
different  values of the mass parameter $\Theta$
of  Eq.~(\ref{eq:index}).  The
pseudorapidity  of  both  quark  jets  is fixed at $\eta  = 0$, and
the  transverse  momentum  of the soft gluon is $k_T =
10$~GeV.  In (d) we show  the  threshold  result  $\Theta=1$
($E_q = m_q$). The units of ${\cal R}_{\Theta}$
are GeV$^{-2}$.}
\label{fig:threshold}
\end{figure}

\begin{figure}[p] 
\unitlength1cm 
\begin{center} 
\epsfig{figure=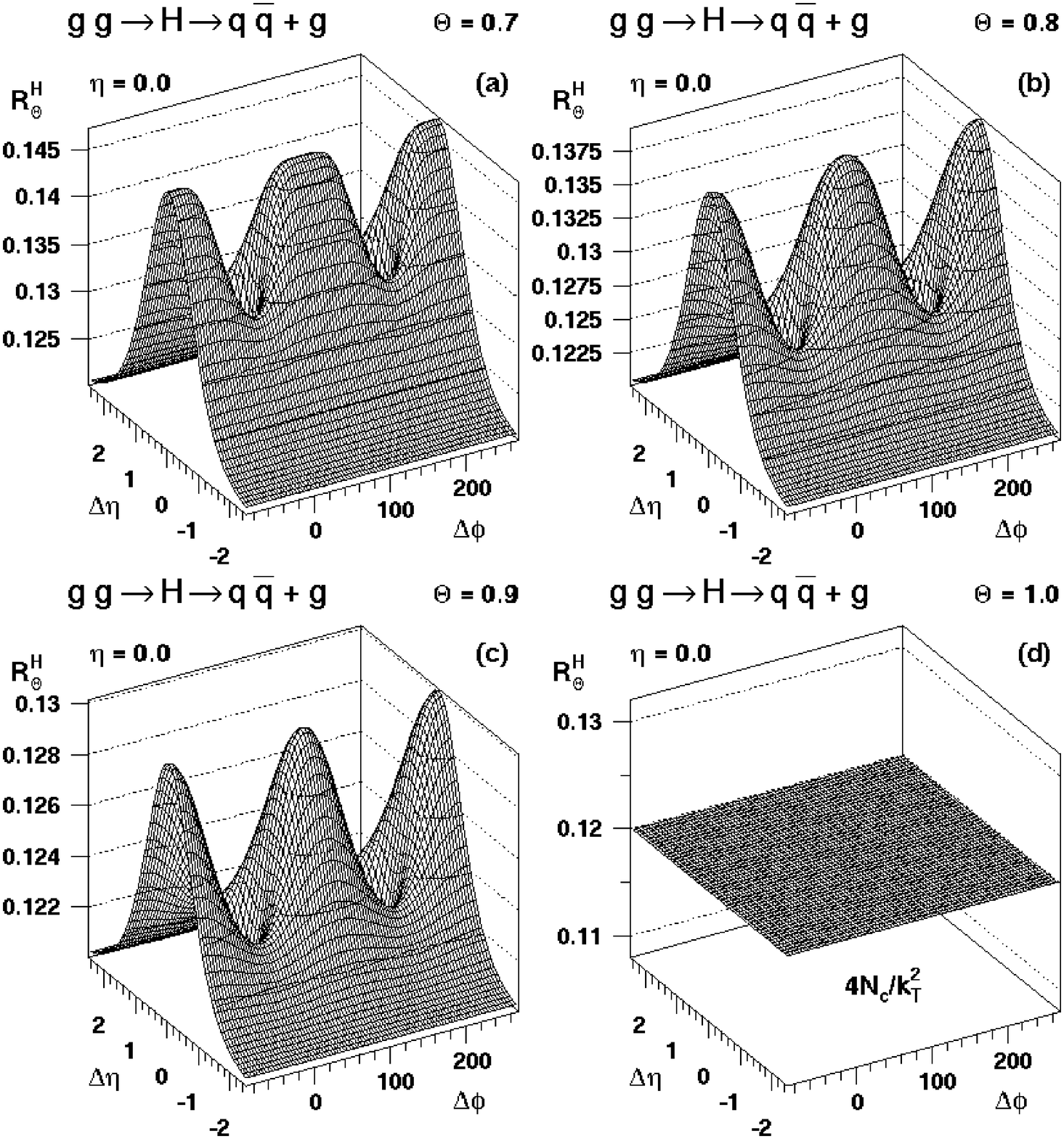,width=\textwidth}
\end{center}
\caption[]{The   antenna   patterns    ${\cal{R}}^{H}_{\Theta}$   of
Eq.~(\ref{massq})  for the process
$gg\rightarrow H(\to q\bar{q})+g$  with
different  values of the mass parameter
$\Theta$ of   Eq.~(\ref{eq:index}).  The
pseudorapidity  of  both  quark  jets  is fixed at $\eta  = 0$, and
the  transverse  momentum  of the soft gluon is $k_T =
10$~GeV.  In (d) we show  the  threshold  result  $\Theta=1$
($E_q = m_q$). The units of ${\cal R}_{\Theta}$
are GeV$^{-2}$.}
\label{fig:thresholdH}
\end{figure}

We first study the   behaviour   of  the radiation patterns
$R^{QCD}_{\Theta}$   and   $R^{H}_{\Theta}$   in  the  threshold
limit  in which  $m_q=E_q=M_H/2$, i.e. $\Theta=1$.
In fact setting   $\eta=0$ we can readily derive
the  general expressions for the antennae  for any value of $\Theta$.
Figures~\ref{fig:threshold} and \ref{fig:thresholdH} show the radiation patterns
for various values of $\Theta$ near and at threshold.
Notice how the strong peaking structure seen in the massless
case (Fig.~\ref{fig:eta0}) disappears as the threshold is approached.
In fact for $\Theta=1$, the patterns do not depend on $\Delta\phi$ at
all. This can be seen from the analytic results. First, for $\Theta=1$
we have $\lbrack 34 \rbrack = \lbrack 33 \rbrack = \lbrack 44  \rbrack$
and so, from  Eq.~(\ref{eq:higgsrat}),
\beq
{\cal{R}}^H_{\Theta=1} = 2N_c\lbrack 12 \rbrack =  \frac{4}{k_T^2}N_c\; ,
\eeq
independent of $\Delta\eta$  and  $\Delta\phi$, see
Fig.~\ref{fig:thresholdH}(d).

For  ${\cal{R}}^{QCD}_{\Theta}$ at  threshold, 
we first note from (\ref{vary}) that $T=U=\mu=\frac{1}{2}$ and thus
\beq \label{eq:xy}
{\cal{X}}_1 = 0, \qquad  {\cal{Y}}_1  =  \frac{2+N_c^2}{2N_c^2(4C_F  -
N_c)}\; .
\eeq
{}From    Eq.~(\ref{massq}) we then have
\begin{eqnarray}
{\cal{R}}^{QCD}_{\Theta = 1}  &=&  2\left(N_c  -C_F + {\cal Y}_1 \right)
\lbrack 12 \rbrack + \left(C_F - {\cal Y}_1 \right) \left\{
\lbrack 13 \rbrack + \lbrack 24  \rbrack +
\lbrack 14 \rbrack + \lbrack 23 \rbrack  \right\}
+ 2( {\cal Y}_1  - C_F ) \lbrack 34  \rbrack
\nonumber \\
& = & \frac{2}{k_T^2}\left( 2N_c  
 - \frac{C_F - {\cal Y}_1 }{\cosh^2(\Delta\eta)  } \right) \; ,   
\end{eqnarray}
which depends on $\Delta\eta$ but not on $\Delta\phi$.
For $|\Delta\eta| \to \infty$ ${\cal{R}}^{QCD}_{\Theta=1}$
approaches the constant value
\beq \label{eq:rqcdspec}
\lim_{|\Delta\eta|\rightarrow\infty}  {\cal{R}}^{QCD}_{\Theta  =  1} =
2N_c\lbrack 12 \rbrack = \frac{4}{k_T^2}N_c\; ,
\eeq
and becomes equal to ${\cal{R}}^{H}_{\Theta=1}$, as in the massless
case. We also see from  Fig.~\ref{fig:threshold}(d)
that ${\cal{R}}^{QCD}_{\Theta=1}$ has an absolute  minimum
at  $\Delta\eta=0$,
\beq \label{eq:rqcdmin}
{\cal{R}}^{QCD}_{\Theta=1}(\Delta\eta=0)   = \frac{2}{k_T^2} 
\left(  2N_c  -  C_F  +
{\cal{Y}}_1\right)   =  \frac{N_c}  {k_T^2} \; \frac{3N_c^2  -
4}{N_c^2-2}\; ,
\eeq
which is numerically $18\%$ lower than the large $\Delta\eta$
value.  Note       the       singularity       in
Eq.~(\ref{eq:rqcdmin}) for $N_c = \sqrt{2}$.

\begin{figure}[phtb] 
\unitlength1cm 
\begin{center} 
\epsfig{figure=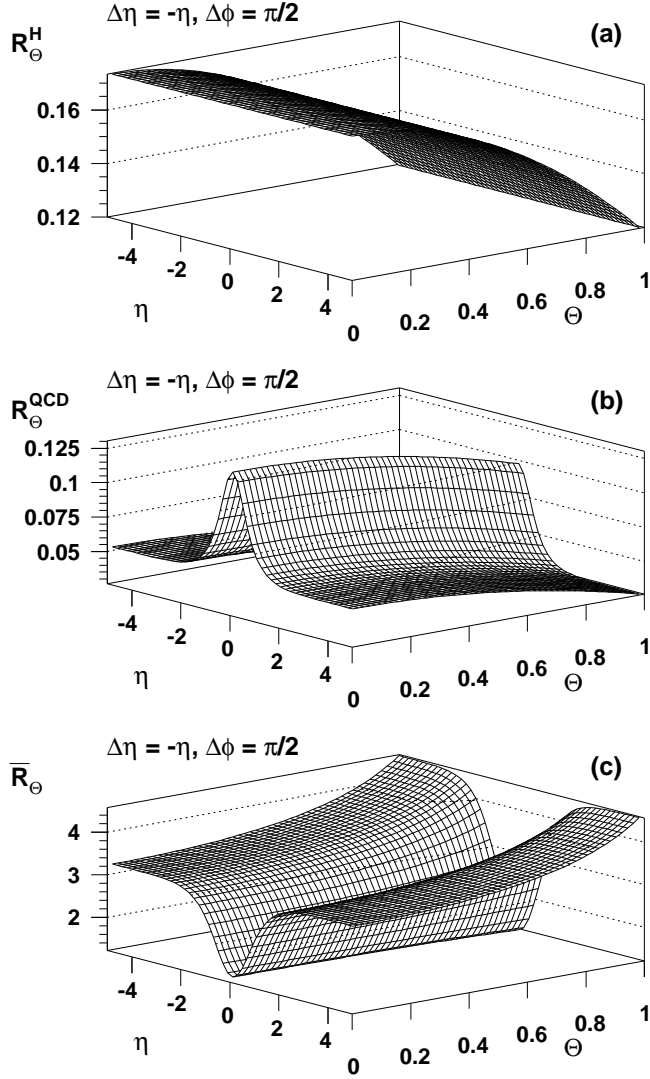,width=10cm}
\end{center}
\caption[]{The antenna patterns for ${\cal{R}}^{H}_{\Theta}$,
${\cal{R}}^{QCD}_{\Theta}$ (in units of GeV$^{-2}$) 
and ${\overline{\cal{R}}}_{\Theta}$ at
the symmetric interjet point
${\cal{P}}_{\rm c}$, for different values of the quark jet rapidity $\eta$
and the mass parameter $\Theta$. The
soft gluon transverse momentum is taken to be $k_T = 10$~GeV.}
\label{fig:newflow}
\end{figure}

We next consider the patterns for arbitrary  $\eta$ and $\Theta$.  With
the  exception  of $\lbrack 12 \rbrack$  all  antennae exhibit an $\eta$
dependence.  We  are again  especially   interested   in   the   value   of
${\cal{R}}^{H}_{\Theta}$ and  ${\cal{R}}^{QCD}_{\Theta}$ at the
symmetric point between the two jets
at   ${\cal{P}}_{\rm  c}  =  (\Delta\eta=   -\eta,
\Delta\phi=  \pi/2)$, as the massless study suggests that at this point
the differences between the signal and background radiation patterns
should be maximal.
When evaluated at  ${\cal{P}}_{\rm  c}$, only $\lbrack 13 \rbrack$,  $\lbrack
14 \rbrack$,  $\lbrack 23 \rbrack$  and  $\lbrack 24 \rbrack$ have an
explicit $\eta$ dependence ($\sim \tanh(\eta)$), whereas
\begin{eqnarray}
\lbrack 12 \rbrack &=&  \frac{2}{k_T^2}, \\ \nonumber
\lbrack 34 \rbrack &=&  \frac{2-\Theta^2}{k_T^2}, \\ \nonumber
\lbrack 33 \rbrack &=& \lbrack 44 \rbrack
=  \frac{\Theta^2}{k_T^2}\; .
\end{eqnarray}
All antennae that are $\eta$ dependent exhibit
an  absolute  maximum  at   ${\cal{P}}_{\rm   c}$  of  $2/k_T^2$  for
$\eta\rightarrow -\infty$ ($\lbrack 13 \rbrack$, $\lbrack 24 \rbrack$)
or for  $\eta\rightarrow\infty$  ($\lbrack  14  \rbrack$, $\lbrack  23
\rbrack$) and vanish for $\eta\rightarrow \pm\infty$ accordingly.  The
fact that there is no $\eta$  dependence  at  ${\cal{P}}_{\rm  c}$ for
$\lbrack 12 \rbrack$,  $\lbrack 34 \rbrack$,  $\lbrack 33 \rbrack$ and
$\lbrack   44    \rbrack$    immediately    yields (see Eq.~(\ref{massh}))
\beq \label{eq:higgscons}
{\cal{R}}^H_{\Theta}({\cal{P}}_{\rm   c}) =
\frac{4}{k_T^2} \left( N_c + C_F ( 1-\Theta^2 ) \right)
\eeq
for all $\eta$, i.e. the  radiation between  the two jets is
completely independent of their separation in rapidity.
This is illustrated in  Fig.~\ref{fig:newflow}(a).
Note that the massless result (Eq.~(\ref{eq:higgstriv}))
is reproduced for $\Theta=0$. The corresponding expression
for ${\cal{R}}^{QCD}_{\Theta}$ {\em is} $\eta$ dependent and
reads
\beq \label{eq:qcdcons}
{\cal{R}}^{QCD}_{\Theta}({\cal{P}}_{\rm   c})  =
\frac{1}{k_T^2}      \frac{\left\{    4(1-\Theta^2)     +2N_c^2     (
N_c^2-2)(2-\Theta^2)      \right\}    \cosh^2(\eta)      +      N_c^2
\left\{ 2-\Theta^2(2-N_c^2)\right\} }{N_c^2( 4C_F\cosh^2(\eta)- N_c)}\; .
\eeq
For   fixed   $\Theta$,
${\cal{R}}^{QCD}_{\Theta}({\cal{P}}_{\rm   c})$
always  shows  an  absolute  maximum  for
$\eta=0$ (see Fig.~\ref{fig:newflow}(b))
with a  $\Theta$  dependence  which  again  is  maximal for
the massless case $\Theta=0$,
with the value  given in  Eq.~(\ref{eq:qcd00pi2}).
Once again defining the ratio of
signal to background radiation patterns as
$\overline{\cal{R}}_{\Theta} =
{\cal{R}}^H_{\Theta} /{\cal{R}}^{QCD}_{\Theta}$,
we see that $\overline{\cal{R}}_{\Theta}$
has a local maximum at ${\cal{P}}_{\rm c}$, the value of which
depends on $\eta$ and $\Theta$, see Fig.~\ref{fig:newflow}(c).
The value at $\eta=0$ is
\beq \label{eq:maxrr}
\overline{\cal{R}}_{\Theta}
({\cal{P}}_{\rm c},\eta=0) = \frac{4 N_c^2 (4C_F-N_c) (C_F(\Theta^2-1)-N_c)}
{(N_c^4+4)(\Theta^2-1) + 2N_c^2(3-\Theta^2)}\; ,
\eeq
which  actually shows a very weak $\Theta$  dependence.  It is maximal
for massless quarks ($\Theta=0$)
with the value ($=1.3285$) already given in Eq.~(\ref{eq:maxrat0}),
and is minimal for $\Theta=1$ with the value
\beq
\overline{\cal{R}}_{\Theta=1}({\cal{P}}_{\rm c},\eta=0) =
 4\frac{N_c^2-2}{3N_c^2 - 4} = 1.2174\; .
\label{eq:dccentre}
\eeq

For the massless case,  $\overline{\cal{R}}({\cal{P}}_{\rm c})$
increased with increasing jet separation (i.e. increasing $\eta$).
This is again true for the massive case, as shown in
 Fig.~\ref{fig:newflow}(c). In the limit $|\eta| \to \infty$ we find
\beq
\lim_{|\eta|\rightarrow\infty}
\overline{\cal{R}}_{\Theta}({\cal{P}}_{\rm c},\eta) =
\frac{4 \left( N_c + C_F ( 1-\Theta^2 ) \right) N_c(N_c^2 - 1)}
{2(1-\Theta^2) + N_c^2 (N_c^2 - 2)(2-\Theta^2)}\; ,
\eeq
which is a monotonically increasing function of $\Theta$.
The values at $\Theta = 0,1$ are $3.25,4.57$ respectively, for $N_c =
3$.

In summary, the relative difference between the radiation patterns
for the Higgs signal  and QCD background processes is maximal
at the symmetric interjet point, as depicted in Fig.~\ref{fig:event}.
The ratio (signal$/$background) of the radiation patterns at this point
depends on the rapidity of the jets and the quark mass. It is smallest
($\overline{\cal{R}} = 1.33$) for massless, central jets, and largest
for massive, large--rapidity jets ($\overline{\cal{R}} = 4.57 $).

\subsection{Radiation inside the `dead cone'}

A final point concerns the radiation {\it inside} the dead cone of
the final--state (massive) quark jets. In this subsection for simplicity
we will only consider centrally produced jets with $\eta= 0$ --- the
generalisation to forward jet production is entirely straightforward.

First we recall the result for
the Higgs signal process $gg \to H \to q \bar q$ for massless quarks
(see Eq.~(\ref{eq:higgsrat0})),
\beq \label{eq:higgsrat1}
{\cal{R}}^H|_{\eta=0}   =  \frac{4}{k_T^2}   \left( N_c
 + \frac{C_F}{\cosh^2(\Delta\eta)
- \cos^2(\Delta\phi)} \right)\; .
\eeq
The second term is singular at the jet centre, $\Delta\eta,\Delta\phi \to
0$, whereas the first term represents a constant `pedestal' 
of radiation from emission off the incoming gluons.
In the massive case ($\Theta > 0$), however, the singularity is removed
and in fact the net contribution to  
the radiation pattern from the combination  $ C_F(2 \lbrack 34 \rbrack
 - \lbrack 33 \rbrack - \lbrack 44 \rbrack)$ vanishes
at the jet centre ${\cal{P}}_{\rm  dc} =
(\Delta\eta= \Delta\phi=0)$, hence
\beq
{\cal{R}}^{H}_{\Theta} ({\cal{P}}_{\rm  dc},\eta=0) = \frac{4}{k_T^2}N_c\; .
\label{eq:dc1}
\eeq
The corresponding result for the QCD
background radiation pattern inside the
dead cone is straightforward to calculate from the
results already presented. We find, again for $\eta = 0$,
\beq
{\cal{R}}^{QCD}_{\Theta}({\cal{P}}_{\rm  dc},\eta=0) =
\frac{N_c}{k_T^2} \frac{3N_c^2-4}{N_c^2-2}\; .
\label{eq:dc2}
\eeq
Interestingly, the results (\ref{eq:dc1},\ref{eq:dc2})
are independent of the quark mass, provided of course that $m_q > 0$.
The effect can be seen in Figs.~\ref{fig:threshold} and
\ref{fig:thresholdH}, where the
value of the radiation patterns at their minima (i.e. inside the dead cones
of the quark jets) is the same for all $\Theta$. The signal to
background ratio in the dead cone is therefore equal to the
value obtained at threshold and given
already in Eq.~(\ref{eq:dccentre}).

\section{Associated Higgs Production}

Higgs production in association with a W boson
$q \bar{q}' \to W^{*}  \to W H$ is a potentially important
discovery channel at both the Tevatron and LHC colliders,
especially for the `intermediate mass' Higgs. The non--hadronic
final state $WH \to \ell \nu_{\ell} \gamma\gamma$ should be relatively
easy to distinguish, but unfortunately has a very small branching ratio,
see for example the recent study in Ref.~\cite{Kun96}.
This raises the question as to whether  a search in the decay channel
\beq \label{eq:aswh}
q \bar{q}' \longrightarrow W^{*} 
\longrightarrow W (\rightarrow \ell{\nu}_{\ell})
 H (\rightarrow b\bar{b})\; ,
\eeq
might be feasible, especially with flavour tagging of both
final--state $b$ quarks \cite{Sta94}.  
Now there is a potentially large irreducible 
background from the QCD process
\beq \label{eq:aswhback}
q \bar{q}'  
\longrightarrow W (\rightarrow \ell\nu_{\ell})
 + b\bar{b}\; ,
\eeq
when $M_{b \bar b} \sim M_H$. 
\begin{figure}[thb] 
\unitlength1cm 
\begin{center} 
\epsfig{figure=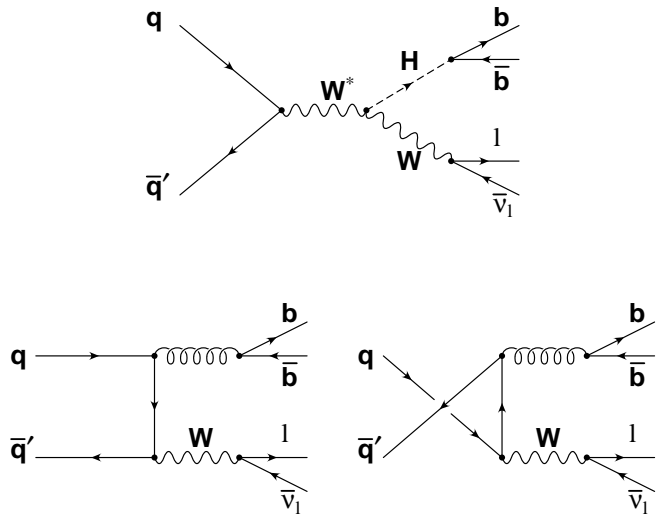,width=9cm}
\end{center}
\caption[]{Feynman graphs for the process $q \bar{q}' 
\rightarrow W^{*} 
\rightarrow W (\rightarrow \ell\bar{\nu}_{\ell})
H (\rightarrow q\bar{q})$ (associated Higgs production) and
the  background process $q \bar{q}' 
\rightarrow W (\rightarrow \ell\bar{\nu}_{\ell})
g^* (\rightarrow q\bar{q})$.}
\label{fig:assokin}
\end{figure}
The signal and background  processes  are illustrated
in Fig.~\ref{fig:assokin}. 

We wish to study the radiation patterns for the processes
(\ref{eq:aswh}) and (\ref{eq:aswhback}), in analogy with the
$gg \to (H \to) b \bar b$ study of the previous section.
We first notice that the colour flows are  exactly
the same as those for the $2\to 2$ scattering processes
$q \bar q \to H \to b \bar b$ and $q \bar q \to g^{*}  \to b \bar b$ \cite{Kho94}.
We can therefore immediately write down the antenna patterns of the
soft gluon radiation:
\beq \label{eq:higgspatt}
{\cal{R}}^{WH}_{\Theta} = 2C_F \left\{ 
\lbrack 12 \rbrack + \lbrack 34 \rbrack \right\} - C_F \lbrack 33 \rbrack
- C_F \lbrack 44 \rbrack\; .
\eeq
\beq \label{eq:qcdpatt}
{\cal{R}}^{Wg}_{\Theta} = \frac{1}{N_c} \lbrack 14;23 \rbrack +
2C_F \left\{ 
\lbrack 13 \rbrack + \lbrack 24 \rbrack \right\} - C_F \lbrack 33 \rbrack
- C_F \lbrack 44 \rbrack\; ,
\eeq
with the momenta labelled as $q(p_1) + \bar{q}'(p_2) \to W + b(p_3) +
\bar{b}(p_4)$ and
$\lbrack 14;23 \rbrack$ defined in Eq.~(\ref{eq:deikonal}).
Note that the Higgs pattern is the same as for $gg\to H \to b \bar b$
apart from colour factor replacement $ N_c \to C_F$ for the 
initial--state $ \lbrack 12 \rbrack $ antenna.

\begin{figure}[thb] 
\unitlength1cm 
\begin{center} 
\epsfig{figure=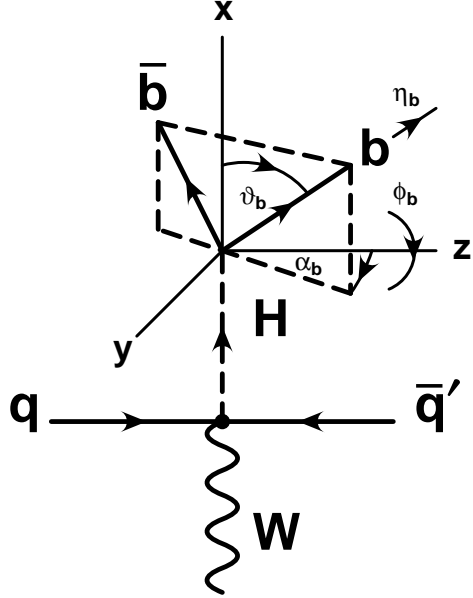,width=7cm}
\end{center}
\caption[]{The kinematics for 
back--to--back Higgs($\to b \bar b$)--$W$ production.
The variables are defined in Eq.~(\ref{eq:whkin}).}
\label{fig:back2back}
\end{figure}

\begin{figure}[thb] 
\unitlength1cm 
\begin{center} 
\epsfig{figure=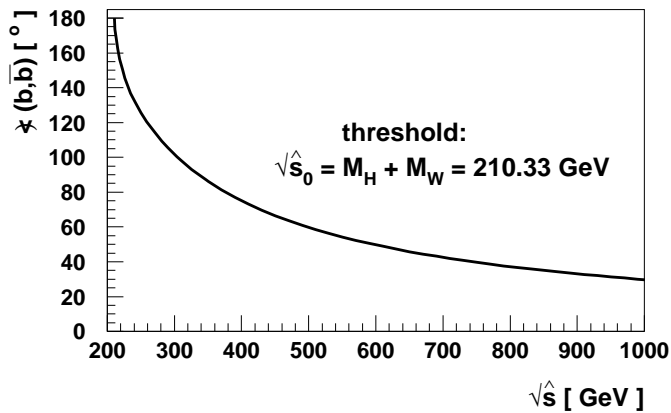,width=10cm}
\end{center}
\caption[]{The opening angle of the $b\bar{b}$ quark pair as
a function of the partonic subprocess energy $\sqrt{\hat{s}}$.}
\label{fig:lhc}
\end{figure}

In order to illustrate the quantitative differences between these
radiation patterns it is necessary to define appropriate kinematics.
Since the leading order processes are now effectively
three--body final states, it is convenient to make some
simplifying assumptions. Thus we assume that the $H$ and the $W$
are produced with zero rapidity, and that the $b$ and $\bar b$ quarks have
equal energy and have polar and azimuthal angles $\vartheta_b$ and $\alpha_b$ with 
respect to the $H$ direction. This configuration is illustrated 
in Fig.~\ref{fig:back2back} and corresponds to the four momenta
\begin{eqnarray}
p_1^{\mu}        & = & (\sqrt{\hat{s}}/2,0,0,\sqrt{\hat{s}}) \nonumber \\
p_{2}^\mu & = & (\sqrt{\hat{s}}/2,0,0,-\sqrt{\hat{s}})\nonumber \\
p_H^{\mu}        & = & (E_H ,p_{TH},0,0)\nonumber \\
p_W^{\mu}        & = & (E_W ,-p_{TH},0,0)\nonumber \\
p_3^{\mu}        & = & (E_b, p_b \cos(\vartheta_b),
                             p_b \sin(\vartheta_b)\sin(\alpha_b),
                             p_b \sin(\vartheta_b) \cos(\alpha_b))\nonumber \\
p_{4}^{\mu}& = & (E_b, p_{TH}-p_b \cos(\vartheta_b),
                            -p_b \sin(\vartheta_b)\sin(\alpha_b),
                            -p_b \sin(\vartheta_b) \cos(\alpha_b))\; .
\label{eq:whkin}
\end{eqnarray}
Conservation of energy and momentum gives
\beq
E_H  = 2E_b = \frac{\hat{s} + M_H^2 - M_W^2}{2\sqrt{\hat{s}}}, \quad
p_{TH} = \sqrt{E_H^2 - M_H^2}, \quad   p_b = \sqrt{E_b^2 - m_b^2}, \quad
\cos(\vartheta_b) = \frac{p_{TH}}{2p_b}\; .
\eeq
The pseudorapidities and azimuthal angles of the $b$ and $\bar b$
quarks are readily found to be
\beq
\tan(\phi_{b,\bar{b}}) = \frac{ p_{(b,\bar{b})y} }{ p_{(b,\bar{b})x}}
= \tan(\vartheta_{b,\bar{b}})\sin(\alpha_{b,\bar{b}})\; ,
\eeq
such that $\alpha_{b,\bar{b}} = \frac{\pi}{2}$ corresponds to $\phi_{b,\bar{b}}
= \vartheta_{b,\bar{b}}$, and 
\beq
\eta_{b,\bar{b}} = \frac{1}{2} \ln\left( \frac{ E_b + p_{(b,\bar{b})z} }
{ E_b - p_{(b,\bar{b})z} } \right)\; .
\eeq
 The soft gluon momentum is defined relative to the $b$--quark jet:
\beq
k^{\mu} = ( k_T \cosh(\eta_b + \Delta\eta), 
k_T\cos(\phi_b + \Delta\phi),
k_T\sin(\phi_b + \Delta\phi),
k_T \sinh(\eta_b + \Delta\eta))\; .
\eeq
Note that the opening angle ($2\vartheta_b$)
between the two $b$ quarks is a function of the 
partonic subprocess energy $\sqrt{\hat{s}}$. The dependence
is illustrated in Fig.~\ref{fig:lhc}.
Note that at threshold ($\sqrt{\hat{s}} =M_W + M_H$)
$2\vartheta_b = 180^\circ$.

Let us now study the radiation patterns in more detail. We assume parameter
values of  $M_H=130$~GeV, $m_b = 4.3$~GeV and $M_W = 80.33$~GeV, and we again fix
the transverse momentum of the soft gluon
to be $k_T = 10$~GeV. The first thing to note is  that for the symmetric
configuration defined above, the radiation pattern for the signal
process is independent of 
the azimuthal angle $\alpha_b$. This follows from the 
absence of antenna involving both initial-- and final--state quarks
in (\ref{eq:higgspatt}). In contrast, there is no such symmetry 
for the background process (\ref{eq:qcdpatt}).

\begin{figure}[thb] 
\unitlength1cm 
\begin{center} 
\epsfig{figure=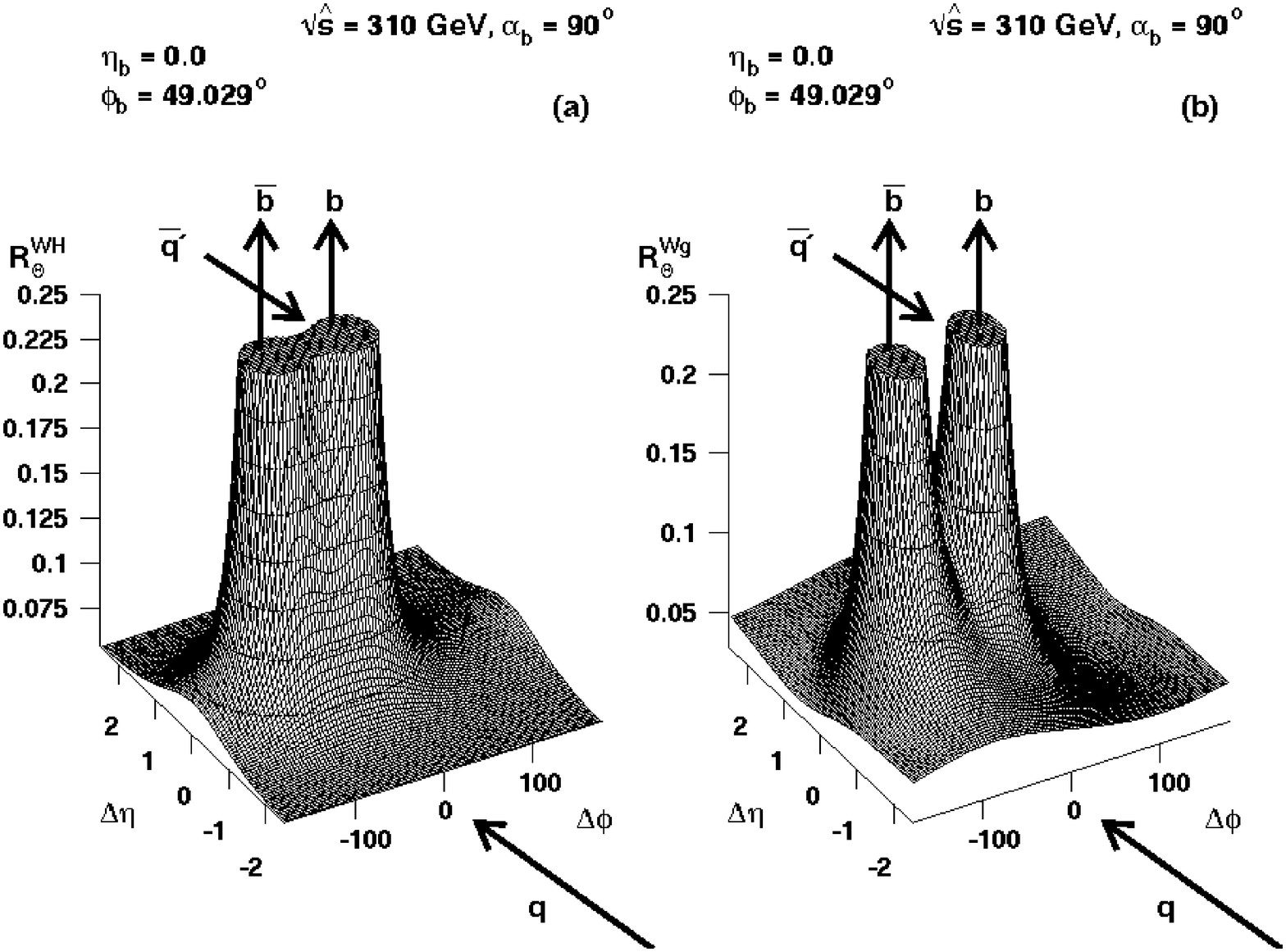,width=\textwidth}
\end{center}
\caption[]{The antenna patterns (in units of GeV$^{-2}$)
for the signal ${\cal{R}}^{WH}_{\Theta}$
(Eq.~(\ref{eq:higgspatt})) and the background ${\cal{R}}^{Wg}_{\Theta}$
(Eq.~(\ref{eq:qcdpatt})) for associated Higgs production
at subprocess centre--of--mass energy  $\sqrt{\hat{s}}= 310$~GeV.
The directions of the incoming quarks $q$ and $\bar{q}'$  and of the $b$ and
$\bar{b}$ quarks are indicated.}
\label{fig:310gev}
\end{figure}

\begin{figure}[thb] 
\unitlength1cm 
\begin{center} 
\epsfig{figure=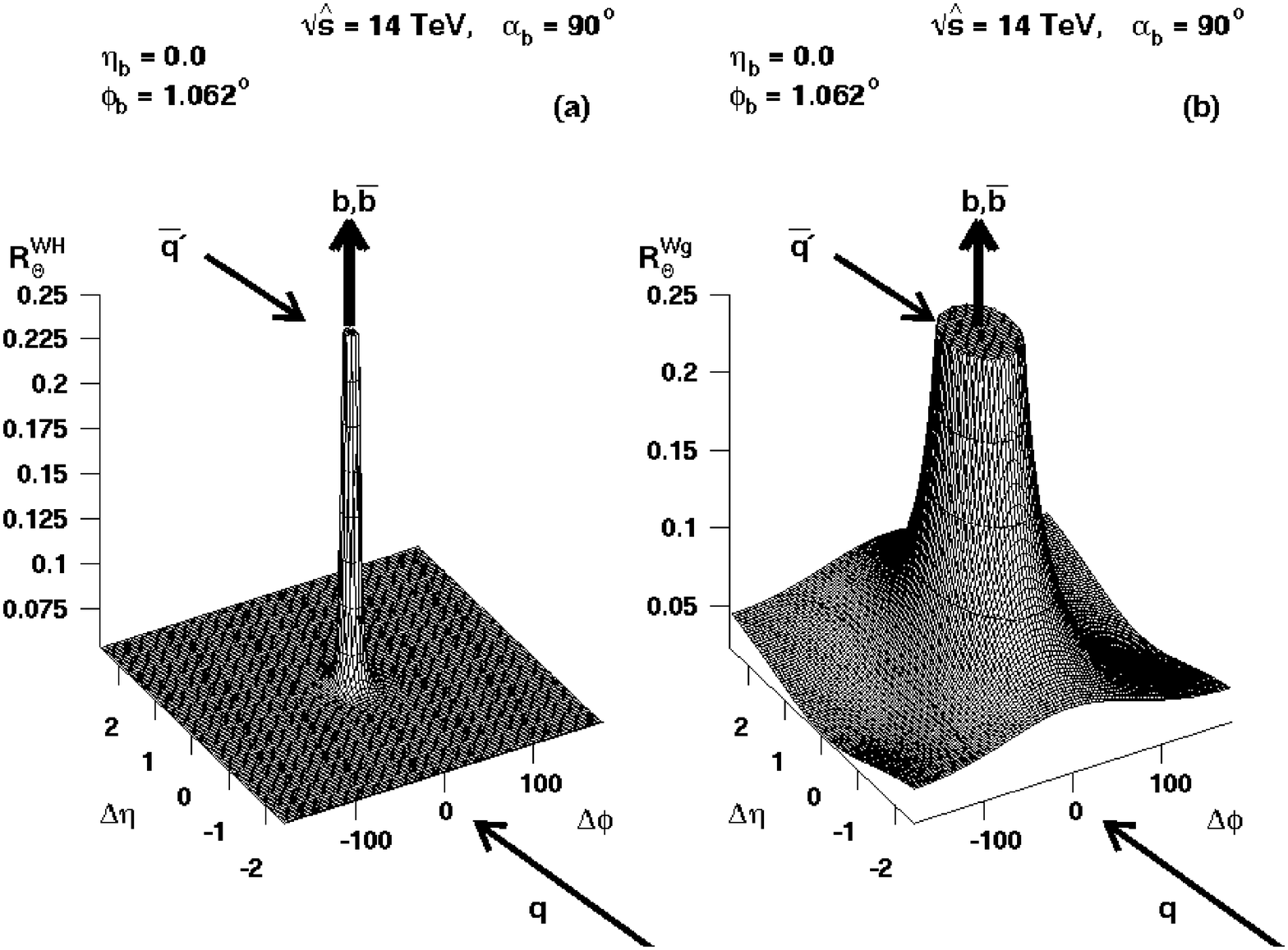,width=\textwidth}
\end{center}
\caption[]{Same as Fig.~\ref{fig:310gev} but now for a subprocess
centre--of--mass energy of $\sqrt{\hat{s}}= 14$~TeV.}
\label{fig:14tev}
\end{figure}

A more  striking difference is seen if
we vary $\sqrt{\hat{s}}$. According to Fig.~\ref{fig:lhc} the angle
between the final--state quarks decreases with increasing
$\sqrt{\hat{s}}$ with the effect that the two quark jets eventually
merge for large centre--of--mass energies. 
 Figs.~\ref{fig:310gev} and 
\ref{fig:14tev}  show the signal (\ref{eq:higgspatt}) and 
background (\ref{eq:qcdpatt}) radiation patterns for the average
value ($\sqrt{\hat{s}}= 310$~GeV) and for an extreme value
($\sqrt{\hat{s}}= 14$~TeV) respectively.\footnote{Notice that 
at threshold, $\sqrt{\hat{s}}_0 = M_H + M_W$, the $b$ and $\bar b$
are produced back--to--back, and the discussion is almost identical
to the direct production case studied earlier, apart from colour
factor differences arising from having incoming quarks instead of gluons.}
 The azimuthal angle $\alpha_b$
is fixed at $90^\circ$ in both cases.
For $\sqrt{\hat{s}} = 310$~GeV the opening angle between the $b$ and the
$\bar{b}$ quarks is  approximately $ 100^{\circ}$. As $\alpha_b= 90^{\circ}$ the $b-\bar{b}$
plane is orthogonal to the $q\bar{q}'-WH$ scattering plane
 (see Fig.~\ref{fig:back2back})
and thus $\eta_b = \eta_{\bar{b}} = 0$. We see immediately that the main feature of our
direct production study described earlier still holds. The most striking difference between 
the signal ${\cal{R}}^{WH}_{\Theta}$ and the background 
${\cal{R}}^{Wg}_{\Theta}$ is the relative  suppression of radiation {\it between}
the $b$--quark jets for the latter. There is a factor of approximately 2
difference between signal and background radiation in the interjet region, 
in qualitative agreement
with the results obtained for direct Higgs production.
If we now increase the subprocess centre--of--mass energy 
 the two $b$--quark jets merge, forming a narrow colour singlet and octet state
 for the signal and background respectively. The situation for the extreme
 case  $\sqrt{\hat{s}}
= 14$~TeV is shown in Fig.~\ref{fig:14tev}. Notice that
for the signal process the soft gluon radiation becomes 
trapped in a very small tube. Outside the merged jets the radiation pattern
completely flattens out.  In contrast, for the background process
there is still significant radiation between the initial-- and final--state
quark directions. In fact the distribution here is essentially identical
to that for the $ q\bar{q}' \to W g$ process studied in Ref.~\cite{Kho97}.
In other words, the radiation pattern  acts as a `partonometer'
\cite{Ell96}
in measuring the colour charge of the outgoing large $p_T$ partonic
system.

\section{Conclusions}

We have studied the distribution of soft hadrons or jets
accompanying the production and $q \bar q$ decay of light
Higgs bosons at high-energy hadron colliders, and compared the
distributions with those of the irreducible QCD backgrounds. We find
significant differences between the signal and background distributions,
which suggests that the study of the topology of hadron flow in such
events could provide an important additional discriminatory tool. For
example we have shown (see Fig.~\ref{fig:eta0}) that the distribution of soft
hadrons transverse to the scattering plane in centrally produced
$H \to b \bar b$ events is approximately $4/3$ larger than that for
QCD $gg \to b \bar b$ events with the same kinematics, while close
to the beam axis and final--state jet directions the signal and
background distributions are the same. The differences result from the
different colour flow in the two processes.

Although in this paper we have focused on light Higgs bosons with
a dominant $b \bar b$ decay mode, we would like to make some 
additional remarks
concerning heavier Higgs bosons. Consider, for example, the `gold--plated'
$gg \to H \to Z^0 Z^0 \to 4 l ^\pm$ discovery channel for a heavy
($M_H > 2 M_Z$) Standard Model Higgs boson. The dominant
irreducible background comes from the $q \bar q \to Z^0 Z^0$ process.
In the language of Section~\ref{sec:antenna}, the antenna patterns for these
signal and background processes are simply  $2N_c [12]$ and $2C_F [12]$
respectively. In other words, soft hadrons or jets with fixed transverse
momentum should be uniformly distributed in the $(\eta,\phi)$ plane
in both cases,
but  with an enhancement of $9/4$ for the signal relative to the
background. A simple on--/off--resonance comparison should therefore
show a significant difference.

Our results are based on the soft--gluon/LPHD hypothesis \cite{adkt}. 
The success of
this approach has recently received a new quantitative 
confirmation from experiments
at the Tevatron $ p \bar p $ collider. However, it will be important to
extend our work by incorporating a realistic Monte Carlo simulation
which will allow detector effects to be included. We believe that the
results presented in this paper make such an effort very worthwhile.

\subsection*{Acknowledgements}
We thank John Ellis for useful discussions.
This work was supported in part by the EU Fourth Framework Programme `Training
and Mobility of Researchers', Network `Quantum Chromodynamics and the Deep
Structure of Elementary Particles', contract FMRX-CT98-0194 (DG 12 - MIHT).



\begin{thebibliography}{99}
\bibitem{DKT}\label{ref:DKT}  Yu.L.~Dokshitzer, V.A.~Khoze and
S.I.~Troyan, in Proc. 6th Int. Conf. on Physics in Collision,
ed. M.~Derrick (World Scientific, Singapore, 1987), p.417. \\
Yu.L.~Dokshitzer, V.A.~Khoze and
S.I.~Troyan, Sov. J. Nucl. Phys. {\bf 46} (1987) 712.
   
\bibitem{DKMT}\label{ref:DKMT} 
Yu.L.~Dokshitzer, V.A.~Khoze,
A.H.~Mueller and S.I.~Troyan, Rev. Mod. Phys. {\bf 60} (1988) 373.\\
Yu.L.~Dokshitzer, V.A.~Khoze and
S.I.~Troyan in:  Advanced Series on Directions in High Energy
Physics, Perturbative Quantum Chromodynamics, ed.
A.H.~Mueller (World Scientific, Singapore), v. 5 (1989) 241.

\bibitem{book}\label{ref:book} 
Yu.L.~Dokshitzer, V.A.~Khoze,
A.H.~Mueller and S.I.~Troyan, \lq\lq Basics of Perturbative
QCD", ed. J.~Tran Thanh Van, Editions Fronti\'{e}res,
Gif-sur-Yvette, 1991.

\bibitem{emw}\label{ref:emw}
 R.K.~Ellis, G.~Marchesini and
B.R.~Webber, Nucl. Phys. {\bf B286} (1987) 643; Erratum
 Nucl. Phys. {\bf B294} (1987) 1180.\\
R.K.~Ellis, presented at ``Les Rencontres de
Physique de la Vallee d'Aoste'', La Thuile, Italy, March 1987, preprint
FERMILAB-Conf-87/108-T (1987). 

\bibitem{DKTSJNP}\label{ref:DKTSJNP} 
Yu.L.~Dokshitzer, V.A.~Khoze and
S.I.~Troyan, Sov. J. Nucl. Phys. {\bf 50} (1989) 505.

\bibitem{DKS}\label{ref:DKS} 
Yu.L.~Dokshitzer, V.A.~Khoze and T.~Sj\"{o}strand,
 Phys. Lett. {\bf B274} (1992) 116.

\bibitem{MarWeb}\label{ref:MarWeb}
G.~Marchesini and B.R.~Webber, Nucl. Phys. {\bf B330} (1990) 261.

\bibitem{ZEPPEN}\label{ref:ZEPPEN} 
D.~Zeppenfeld, Madison preprint MADPH-95-933 (1996).

\bibitem{Ell96}\label{ref:Ell96}  J.~Ellis,   V.A.~Khoze   and   W.J.~Stirling,
{\it Zeit. Phys.} {\bf  C75}  (1997) 287.

\bibitem{Kho97}\label{ref:Kho97}  V.A.~Khoze and W.J.~Stirling,
{\it Zeit. Phys.} {\bf  C76}  (1997) 59.

\bibitem{Amu98}\label{ref:Amu98} J.~Amundson, J.~Pumplin and  C.~Schmidt,
{\it  Phys. Rev.} {\bf  D57} (1998) 527.

\bibitem{klo}\label{ref:klo}  V.A.~Khoze, S.~Lupia and W.~Ochs,
preprint CERN-TH/97-199, hep-ph/9711392.

\bibitem{adkt}\label{ref:adkt} 
Ya.I.~Azimov, Yu.L.~Dokshitzer, V.A.~Khoze and S.I.~Troyan,
 Z. Phys. {\bf C27} (1985) 65; {\bf C31} (1986) 213.

\bibitem{CDF}\label{ref:CDF} CDF collaboration:
F.~Abe {\it  et al.}, {\it  Phys. Rev.} {\bf  D50} (1994) 5562.
 
\bibitem{D0}\label{ref:D0} D0 collaboration: B.~Abbott {\it  et al.}, 
{\it Phys. Lett.} {\bf B414} (1997) 419; preprint FERMILAB-Conf-97-372-E;
N.~Varelas, preprint FERMILAB-Conf-97-346-E.


\bibitem{LHC}\label{ref:LHC} Proceedings of the ``{\it Large Hadron Collider Workshop}'', 
Aachen, 4-9 October 1990, eds. G.~Jarlskog and D.~Rein, 
Report CERN 90-10, ECFA 90-133, Geneva, 1990.  

\bibitem{ATLAS}\label{ref:ATLAS} ATLAS Technical Proposal, CERN/LHC/94-43 LHCC/P2 (December 1994).

\bibitem{CMS}\label{ref:CMS} CMS Technical Proposal, CERN/LHC/94-43 LHCC/P1 (December 1994).

\bibitem{Hey97}\label{ref:Hey97}  M.~Heyssler and W.J.~Stirling,
{\it Phys. Lett.} {\bf B407} (1997) 259.

\bibitem{Azi85}\label{ref:Azi85}  Ya.I.~Azimov, Yu.L.~Dokshitzer, V.A.~Khoze and
S.I.~Troyan, {\it Phys.  Lett.}  {\bf B165} (1985) 147.

\bibitem{Kho94}\label{ref:Kho94}  V.A.~Khoze, J.~Ohnemus and W.J.~Stirling, {\it
Phys.  Rev.}  {\bf D49} (1994) 1237.

\bibitem{Dok91a}\label{ref:Dok91a} Yu.L.~Dokshitzer, V.A.~Khoze and S.I.~Troyan,
{\it J.  Phys.}  {\bf G17} (1991) 1481; {\it ibid.}\/ {\bf G17} (1991) 1602;
{\it Phys. Rev.} {\bf D53} (1996) 89.

\bibitem{Kun96}\label{ref:Kun96}  Z.~Kunszt, S.~Moretti and W.J.~Stirling,
{\it Zeit. Phys.} {\bf  C74}  (1997) 479.

\bibitem{Sta94}\label{ref:Sta94} A.~Stange, W.~Marciano and S.~Willenbrock, 
{\it Phys. Rev.} {\bf D50} (1994) 4491.

\end{thebibliography}
\end{document}